\documentclass[iop,twocolappendix]{emulateapj}
\pdfoutput=1
\usepackage{graphicx}        
\usepackage{color}           
\usepackage{breakurl}        
\usepackage{natbib}
\usepackage{epstopdf}
\usepackage{gensymb}
\usepackage{amsmath}
\newcommand{\kms}{km\,s$^{-1}$}

\shorttitle{Plasma diagnostics of an EUV wave}
\shortauthors{Vanninathan et al.}

\begin{document}


\title{Coronal response to an EUV wave from DEM analysis}


\author{K. Vanninathan, A.M. Veronig\altaffilmark{1}, and  K. Dissauer} \affil{IGAM/Institute of Physics, University of Graz, 8010 Graz, Austria} \email{kamalam.vanninathan@uni-graz.at}
\author{M.S. Madjarska} \affil{Armagh Observatory, College Hill, Armagh BT61 9DG, UK}
\and
\author{I.G. Hannah and E.P. Kontar} \affil{SUPA School of Physics and Astronomy, University of Glasgow, Glasgow G12 8QQ, UK}
  \altaffiltext{1}{Kanzelh{\" o}he Observatory/Institute of Physics, University of Graz, 9521 Treffen, Austria}








\begin{abstract}
EUV (Extreme-Ultraviolet) waves are globally propagating disturbances that have been observed since the era of the SoHO/EIT instrument. Although the kinematics of the wave front and secondary wave components have been widely studied, there is not much known about the generation and plasma properties of the wave. In this paper we discuss the effect of an EUV wave on the local plasma as it passes through the corona. We studied the EUV wave, generated during the 2011 February 15 X-class flare/CME event, using Differential Emission Measure diagnostics. We analyzed regions on the path of the EUV wave and investigated the local density and temperature changes. From our study we have quantitatively confirmed previous results that during wave passage the plasma visible in the Atmospheric Imaging Assembly (AIA) 171\,\AA\ channel is getting heated to higher temperatures corresponding to AIA 193\,\AA\ and 211\,\AA\ channels. We have calculated an increase of  6 -- 9\% in density and 5 -- 6\% in temperature during the passage of the EUV wave. We have compared the variation in temperature with the adiabatic relationship and have quantitatively demonstrated the phenomenon of heating due to adiabatic compression at the wave front. However, the cooling phase does not follow adiabatic relaxation but shows slow decay indicating slow energy release being triggered by the wave passage. We have also identified that heating is taking place at the front of the wave pulse rather than at the rear. Our results provide support for the case that the event under study here is a compressive fast-mode wave or a shock.
\end{abstract}


\keywords{Sun: corona, Sun: evolution, waves}



\section{Introduction}
\label{intro}
Globally propagating disturbances in the solar atmosphere have been known since they were first detected by \cite{Moreton1960} in H$\alpha$ images during flare observations, the so-called  Moreton waves. Such phenomena were modeled by \cite{Uchida1968} as the chromospheric imprint of fast mode Magnetohydrodynamic (MHD) waves in the corona. Though, it was not until the launch of the Extreme-ultraviolet Imaging Telescope \citep[EIT,][]{Delaboudiniere1995} on-board the Solar and Heliospheric Observatory \citep[SoHO,][]{Domingo1995} that the proposed coronal counterparts were identified \citep{Moses1997, Thompson1998}. These transient events have since been termed as ``EIT waves'' after the instrument used to discover them. However, due to the debate regarding the true nature of these transients and to incorporate observations made from different instruments, several authors refer to these events by other names such as ``coronal bright fronts'' \citep{Gallagher2011}, ``coronal propagating fronts'' \citep{Schrijver2011}, ``global EUV waves'' \citep{Patsourakos2012}, and ``large-scale coronal propagating fronts'' \citep{Nitta2013}. In this paper we will refer to them with a generic name such as EUV (Extreme-Ultraviolet) waves.

Inconsistencies in speeds between Moreton and EUV waves \citep{Klassen2000}, as well as the observation of stationary fronts \citep{Delannee1999} led to alternate interpretations and models for the observed wave-like signatures. There are two main opposing theories, waves and non-waves, to explain the visible features of EUV waves. In the non-wave models, these transient phenomena are interpreted as ground tracks of successive restructuring of magnetic field lines during the eruption of a coronal mass ejection \citep[CME,][]{Delannee1999, Chen2002, Attrill2007}. The wave models treat them as fast-mode shock waves or large amplitude waves \citep{Thompson1998, Mann1999, Warmuth2001}. A third hybrid model tries to bridge the gap by recognizing the existence of two bright fronts: one consistent with the wave model and the other with the non-wave model \citep{Zhukov2004}. In recent years, this theory has gained support through observations \citep{Liu2010} and MHD simulations \citep{Cohen2009, Downs2012}.

The average speed of EUV waves was measured to be in the range of 200 -- 400\,\kms\ in the EIT data \citep{Thompson2009} and the Solar Terrestrial Relations Observatory (STEREO) data \citep{Muhr2014} and updated to 600\,\kms\ from the Atmospheric Imaging Assembly (AIA) data \citep{Nitta2013}. Fast EUV waves typically decelerate during propagation, which has been interpreted as the nonlinear evolution of large-amplitude fast magnetosonic waves \citep{Vrsnak2006, Long2008, Veronig2008, Warmuth2011, Muhr2014}. Taking advantage of the quadrature configuration of the STEREO satellites, many authors were able to study the 3D structure and evolution of EUV waves \citep{Patsourakos2009, Kienreich2009, Ma2009, Veronig2010, Temmer2011}. It was estimated that most of the EUV wave emission is coming from heights of 80 -- 100\,Mm above the photosphere \citep{Patsourakos2009, Kienreich2009} which is comparable to 1 -- 2 coronal scale heights (distance from the photosphere over which the pressure scales as a factor of 1/e).

\begin{figure*}[ht!]
\includegraphics[width=\textwidth]{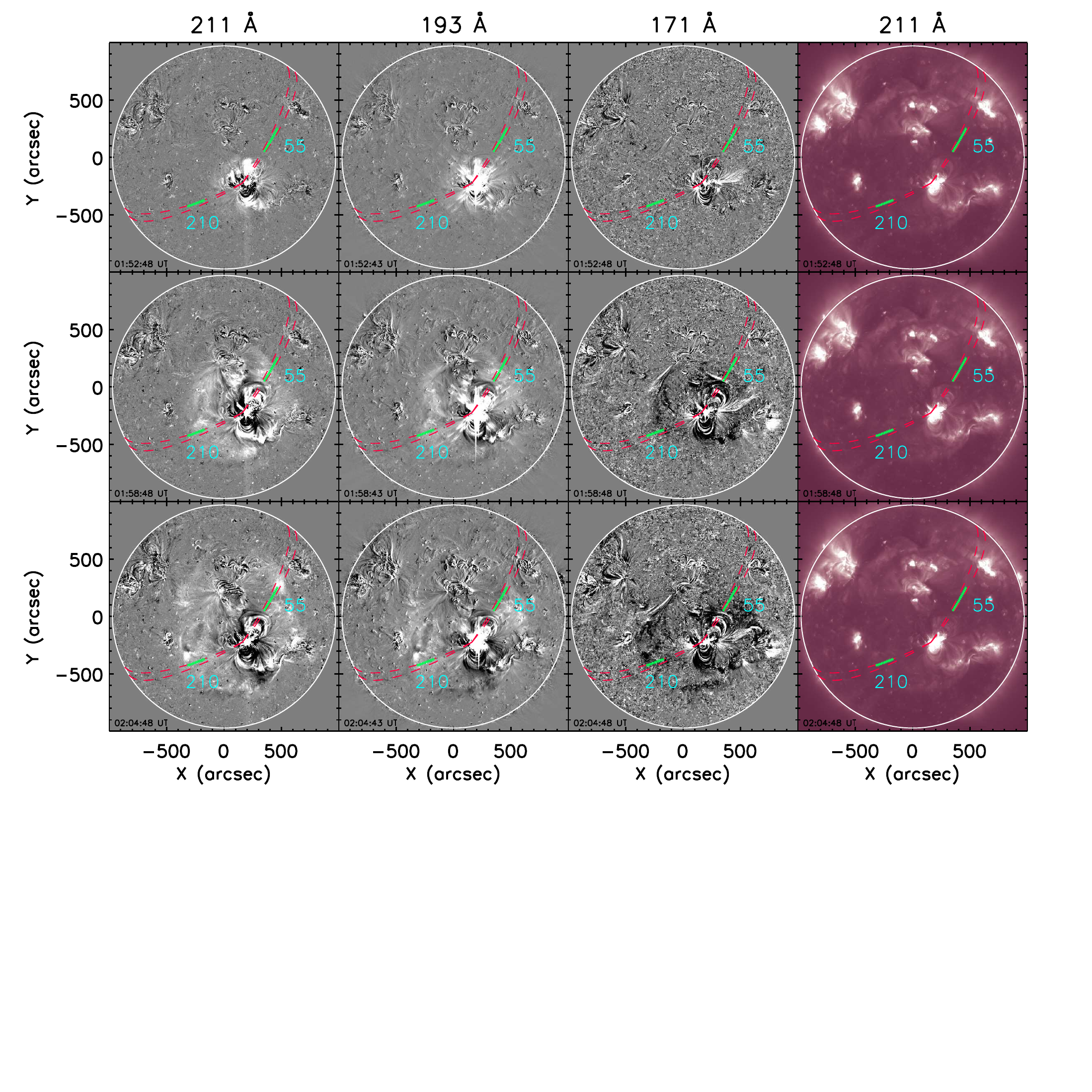}\vspace{-5cm}
\caption{First three columns show base-difference images of the evolution of the wave in 3 EUV channels (211\,\AA, 193\,\AA, 171\,\AA) of AIA in which the wave is clearly visible. Fourth column shows the original data from 211\,\AA\ channel. Co-temporal images are shown in each row. Time is marked at the bottom of each panel. The two sectors along which the EUV wave was studied are marked as 55 and 210, corresponding to the angles they make with the solar west as measured in the counter-clockwise direction, are shown enclosed by the red dashed lines. The green asterisks along each path marks the fixed ROI which was used to further investigate the plasma properties.}
\label{sector}
\end{figure*}

The high cadence and resolution of AIA has helped in improving the knowledge of kinematics of wave fronts and to identify secondary wave components. With regard to solar atmospheric research and plasma physics in general, it is hugely important to study how these waves affect the local plasma and pre-existing magnetic structures as they propagate through. Oscillations and deflections of filaments \citep{Okamoto2004,Shen2014}, coronal loops \citep{Wills-davey1999}, coronal cavities \citep{Liu2012}, and streamers \citep{Tripathi2007} have often been attributed to the passage of EUV waves. It has also been suggested that jets \citep{Shen2014} and sympathetic eruptions \citep{Schrijver2013} can be triggered by such waves.

From observations made with the EIT, the EUV Imager (EUVI) on-board the twin spacecrafts STEREO, and the AIA, EUV waves are typically seen as intensity enhancements in the 193\,\AA\ (AIA), the 195\,\AA\ (EIT, EUVI), the 211\,\AA\ (AIA), the 284\,\AA\ (EUVI), and sometimes in the 335\,\AA\ (AIA) channels while the 171\,\AA\ channel (EIT, EUVI, AIA) often shows co-spatial emission decrease. This has prompted many authors to suggest that plasma from the 171\,\AA\ channel is being heated, leading to the ionization of Fe~{\sc ix/x} to higher ionization states and which results in the increased intensity as registered in the hotter channels \citep{Wills-davey1999, Dai2010, Schrijver2011}. The temperature range of the channels in which EUV waves are observed is about 1 -- 2.5\,MK.

Studies on the plasma properties of EUV waves have been limited by the low chance of encountering such transient phenomena in the limited field-of-view (FOV) of spectrometers. \cite{Veronig2011} conducted a dedicated observing campaign for registering EUV waves with the Hinode/EIS (EUV Imaging Spectrometer) in a sit-and-stare mode. The authors were successful in obtaining one example where an EUV wave passed along the FOV of the EIS slit. The results from this campaign showed down-flows of 20\,\kms\ followed by up-flows of about 5\,\kms\ representing compression at and relaxation behind the wave front, respectively \citep{Harra2011, Veronig2011}. Density increase at the wave front, determined from the EIS Fe~{\sc xiii} line pairs was found to be less than 10\% for the event under study \citep{Veronig2011}. For more detailed analysis of the plasma properties for many events we have to rely on Differential Emission Measure (DEM) techniques that can be applied to imaging data. \cite{Kozarev2011} compared the wave front with a pre-event region using DEM analysis and noted a density increase of 12 -- 18\% at the location of the wave front. Their DEMs at the wave front also increased at higher temperatures which is consistent with plasma heating and compression.

In this paper, we use DEM based analysis to study the response of the surrounding corona to the passage of a strong EUV wave associated with the first X-class flare of cycle 24 on 2011 February 15. The paper is organized as follows, in section~\ref{data} we describe our data and methods of analysis. The results are presented in section~\ref{results} and discussed in section~\ref{discussions}. Finally, in section~\ref{conclusion} we give the summary and conclusions of this work. 

\section{Observations and Data analysis}
\label{data}
\subsection{Event overview}
The AIA \citep{Lemen2012} instrument on-board the Solar Dynamics Observatory \citep[SDO,][]{Pesnell2012} images the Sun and its atmosphere using six EUV filters. These filters put together cover a broad temperature range which facilitates the use of these data for the purpose of DEM analysis. In this paper we study the EUV wave associated with one of the strongest flare events that occurred after the launch of SDO/AIA. The 2011 February 15 event is an X2.2 class flare within AR 11158 (NOAA coordinates: S21 W28). This event has been the interest of previous studies related to sun quakes \citep{Kosovichev2011, Zharkov2011}, sunspot rotation \citep{Jiang2012, Wang2014}, changes in the orientation of photospheric magnetic fields \citep{Wang2012, Gosain2012}, CME kinematics \citep{Gopalswamy2012}, CME-CME interaction \citep{Temmer2014, Shanmugaraju2014} and, of particular relevance to the present study, EUV waves \citep{Schrijver2011, Olmedo2012}. 

For this event, \cite{Schrijver2011} performed global magnetic field modeling and suggested that the observed EUV wave-like signatures are a result of compression between expanding coronal loops, which eventually become a CME, and an overlying helmet streamer structure. \cite{Olmedo2012} studied the interaction of the EUV wave with a coronal hole and detected reflected and transmitted wave components that are consistent with a fast mode MHD wave interpretation. 

This paper is related to the EUV wave that was associated with this event. The initiation time of the flare, according to GOES data, was 01:44\,UT and the peak time was 01:56\,UT. By examining the AIA 211\,\AA\ channel images, we found that the first instance when the EUV wave pulse could be distinguished was at 01:50\,UT and it was seen propagating isotropically across the disk. We studied the evolution of the EUV wave along two sectors which are at $55\degree$ and $210\degree$ (indicated by the red dashed lines in Figure~\ref{sector}) as measured in the anti-clockwise direction from the solar west. These particular directions were chosen because it was possible to observe a strong signal of the wave front, in these directions, for long distances unhindered by other coronal structures such as active regions, coronal holes, coronal loops, etc.

\subsection{Perturbation profiles}
As EUV waves cannot be easily visualized directly from original intensity images, images enhanced by a base-difference technique were used to study the absolute changes in the emission. The base image was taken at 01:45\,UT and was subtracted from each subsequent image taken with one minute cadence. Although AIA data have a higher cadence of 12\,s in the EUV channels, for the current study one minute cadence was sufficient. We only chose those images where the exposure times were constant and not affected by the automatic exposure control algorithm of the instrument.

To be able to conveniently track the EUV wave as it propagates away from the CME site we made use of perturbation profiles which are flux vs. distance plots to analyze the flux along the propagating path of the EUV wave. For calculating the perturbation profiles we used the method adopted by \cite{Muhr2011}. The heliospheric coordinates of the flare given in the GOES data were assumed to be the center of the propagating wave. From this position we selected sectors of $5\degree$ angular width in the chosen directions. The propagation of the EUV wave is affected by local coronal structures and varies in different directions on the Sun. This study is related to the local changes in the quiet corona as the EUV wave passes through. Hence, we thought it necessary to choose a small sector so as to not average out all the properties of the EUV wave front. Along the chosen sectors, we defined successive annuli of $1\degree$ radial width between two constantly growing concentric circles from the wave center to study the properties of the EUV wave. The mean of all the pixel values in a given annulus within a sector were plotted as a function of distance from the wave center to the limb. This procedure was repeated for each time step until the EUV wave faded away. The perturbation profiles were constructed from original images as well as the base-difference images.

\subsection{DEM analysis}
A DEM distribution gives information about the plasma distribution as a function of temperature along a given line-of-sight (LOS). For optically thin emission that is in thermal equilibrium
a DEM, $\phi(T)$, is defined as
\begin{equation}
\phi(T) = n_e^2(T)\,\frac{\mathrm{d}h}{\mathrm{d}T},
\end{equation}
where $n_e$ is the electron density at position $h$ along the LOS at temperature $T$. For spectroscopic observations the line intensity, $I_{\lambda}$, is associated with the DEM by the equation
\begin{equation}
I_{\lambda} = \int A(X)\,G_{\lambda}(n_e, T )\,\phi(T)\,dT,
\end{equation}
where $A(X)$ is the elemental abundance with respect to hydrogen and $G_{\lambda}(n_e,T)$ is the contribution function for a given spectral line. Analogous to this, for broad/narrow band filter observations the intensity in a filter, $I_{filter}$, is associated to the DEM by the equation
\begin{equation}
I_{filter} = \int R_{filter}(T)\,\phi(T)\,\mathrm{d}T,
\end{equation}
where $R_{filter}(T)$ is the temperature response function of the instrument. With the help of inversion techniques or forward fitting models, the DEM from multi-channel observations can be derived using this relationship.

\cite{Hannah2012} have developed a regularized inversion method to reconstruct the DEM from the six AIA channels that are sensitive to coronal temperatures.
This method has been tested for various cases and was found to be reasonably robust and computationally fast \citep{DelZanna2013, Plowman2013}. Additionally, their code provides error bars in both vertical (DEM) and horizontal (temperature) directions. However, this method is also with its limitations. \cite{Hannah2013} have cautioned about the DEM solution at regions with large uncertainties ($\geq$ 0.3 dex) in temperature resolution.
The application of this method also shows some spurious high temperature emission which is most likely contributions of cool lines to the AIA 94\,\AA\ and 131\,\AA\ channels that are currently unaccounted for in the atomic models \citep{Young2014}. The phenomenon that we study in this paper is known to be formed within the thermal range of about 1\,MK to 2.5\,MK. The above mentioned caveats are not affecting this temperature range. Hence for the current study we have applied this DEM technique (henceforth referred to as HK code).

 Average counts in each of the six EUV filters of AIA were determined from the perturbation profiles and used as an input for the DEM code. Apart from data numbers (DNs), the DEM code also accepts errors in DNs as an input. To calculate the uncertainties, we used the method outlined in \cite{Yuan2012} for AIA data, which has been modified from \cite{Aschwandan2000} for the use with TRACE data. The square of the total noise ($\sigma_{noise}$) was determined using the formula 
\begin{eqnarray}
\begin{split}
\sigma^2_{noise}(F) =~	& \sigma^2_{photon}(F) + \sigma^2_{readout} + \sigma^2_{digit} \\
				& +\sigma^2_{compress} + \sigma^2_{dark} + \sigma^2_{subtract} \\
				& +\sigma^2_{spikes}(F) + \sigma^2_{resp},
\end{split}
\end{eqnarray}
which corresponds to the sum of the squares of uncertainties, in units of DNs, in photon Poisson noise ($\sigma_{photon}$), electronic readout noise ($\sigma_{readout}$), digitization noise ($\sigma_{digit}$), compression noise ($\sigma_{compress}$), dark current noise ($\sigma_{dark}$), subtraction noise ($\sigma_{subtract}$), noise due to removal of spikes in the images ($\sigma_{spikes}$), and errors in the response functions ($\sigma_{resp}$). The values for the instrumental errors were taken from \cite{Yuan2012} and \cite{Boerner2012}.

\subsection{Mean plasma temperature and density}
For this study, we made use of the HK code and derived the DEMs from the original data to calculate the DEM weighted average temperature ($\bar{T}$) and  plasma density ($\bar{n}$) using the following expressions (defined in \citealt{Cheng2012})

\begin{equation}
\bar{T}=\frac{\int \phi(T)\,T\,\mathrm{d}T}{\int \phi(T)\,\mathrm{d}T},
\label{temp_eq}
\end{equation}
\begin{equation}
\bar{n}=\sqrt{\frac{\int \phi(T)\,\mathrm{d}T}{h}},
\label{dem_eq}
\end{equation}
where $h$ is the distance along the LOS. From quadrature studies, the height of the EUV waves have been established to be between 80 -- 100 Mm \citep{Patsourakos2009, Kienreich2009}. For the calculations here we assume $h=90$\,Mm.

We also made difference DEMs (referred to as $\Delta$DEM henceforth) by subtracting the DEM of the base image (taken at 01:45\,UT) from the DEM at each subsequent time step. This was used to better illustrate the changes in the DEM during the passage of the EUV wave. This can be considered to be similar to background subtracted images or base images for the purpose of image enhancement.




\section{Results}
\label{results}
\begin{figure}[ht]
\vspace{-0.5cm}
\includegraphics[width=8.5cm]{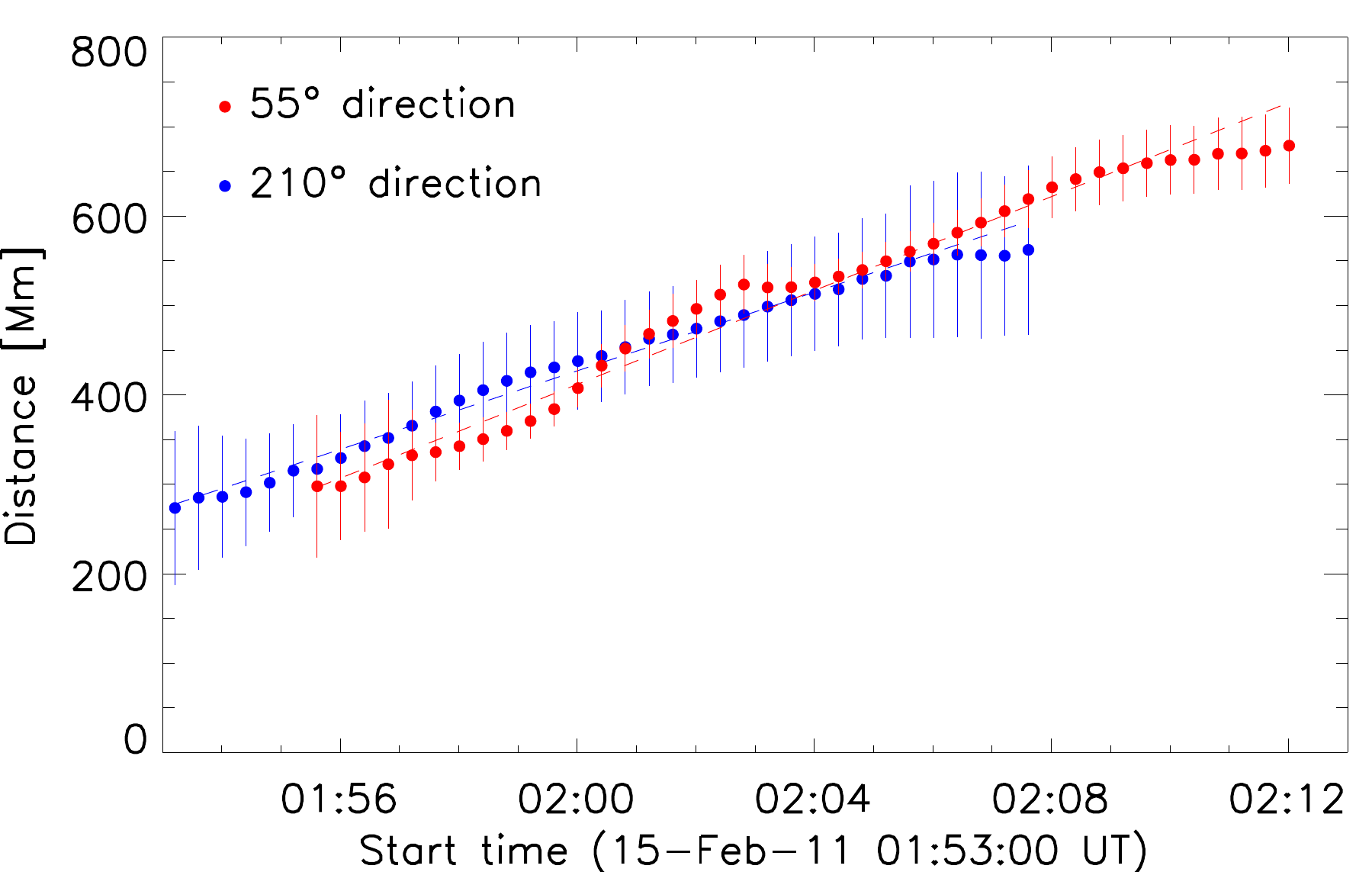}
\caption{Distance vs. time plot along with 1$\sigma$ errors obtained from AIA 211\,\AA\ channel for the propagation of the EUV wave front along 55$\degree$ and 210$\degree$ direction.}
\label{kinematics}
\end{figure}
\begin{figure}[ht]
\includegraphics[width=8.5cm]{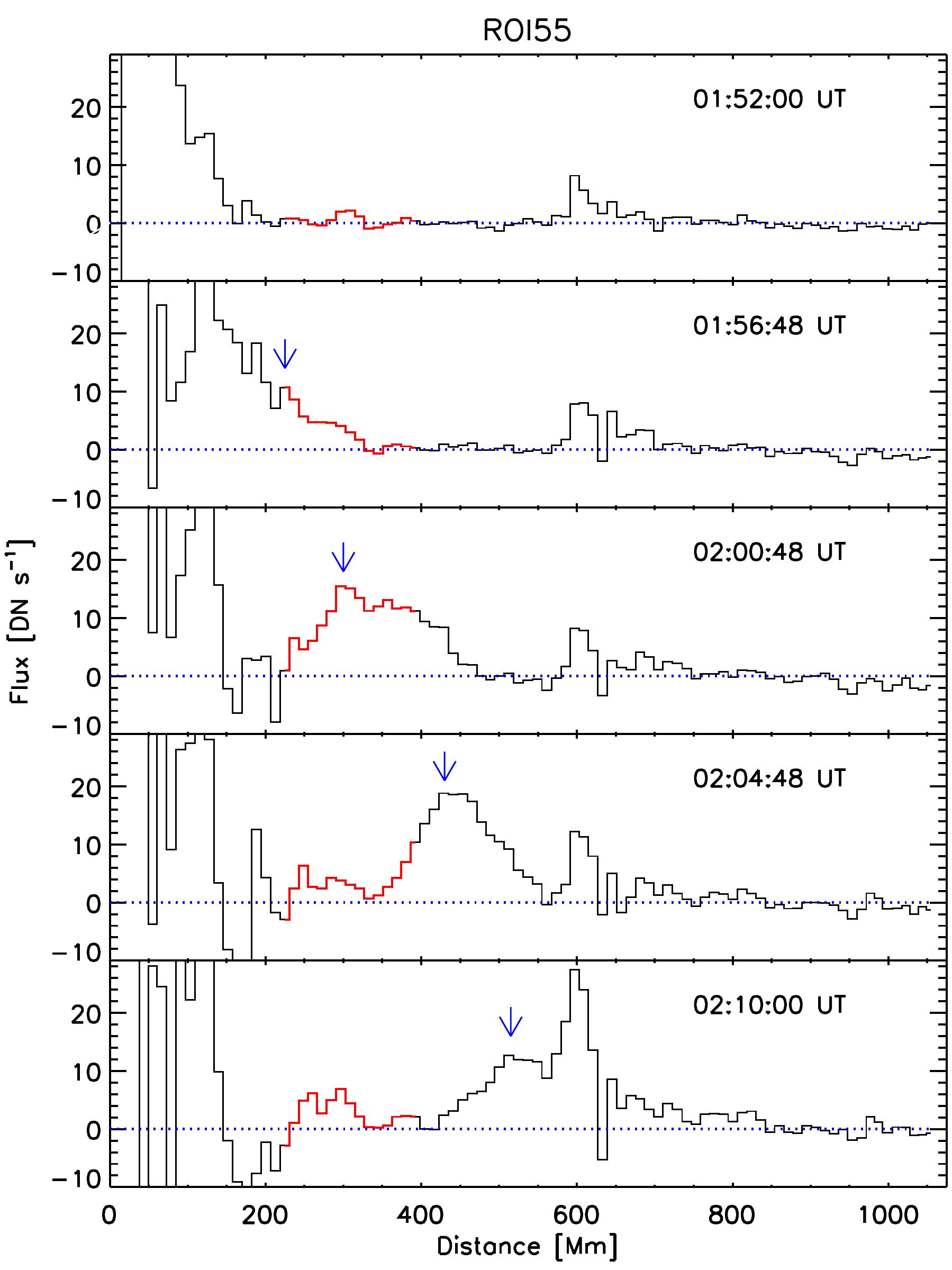}
\caption{Perturbation profiles along the 55$\degree$ direction derived from AIA 211\,\AA\ base-difference images at five different time steps as indicated in the panels. The blue arrow points to the propagating EUV wave pulse. The section of the plots indicated in red corresponds to ROI55 marked by green asterisks along the $55\degree$ direction in Figure~\ref{sector}.}
\label{profiles}
\end{figure}
\begin{figure*}[ht!]
\hspace{0.5cm}
\includegraphics[width=0.45\textwidth]{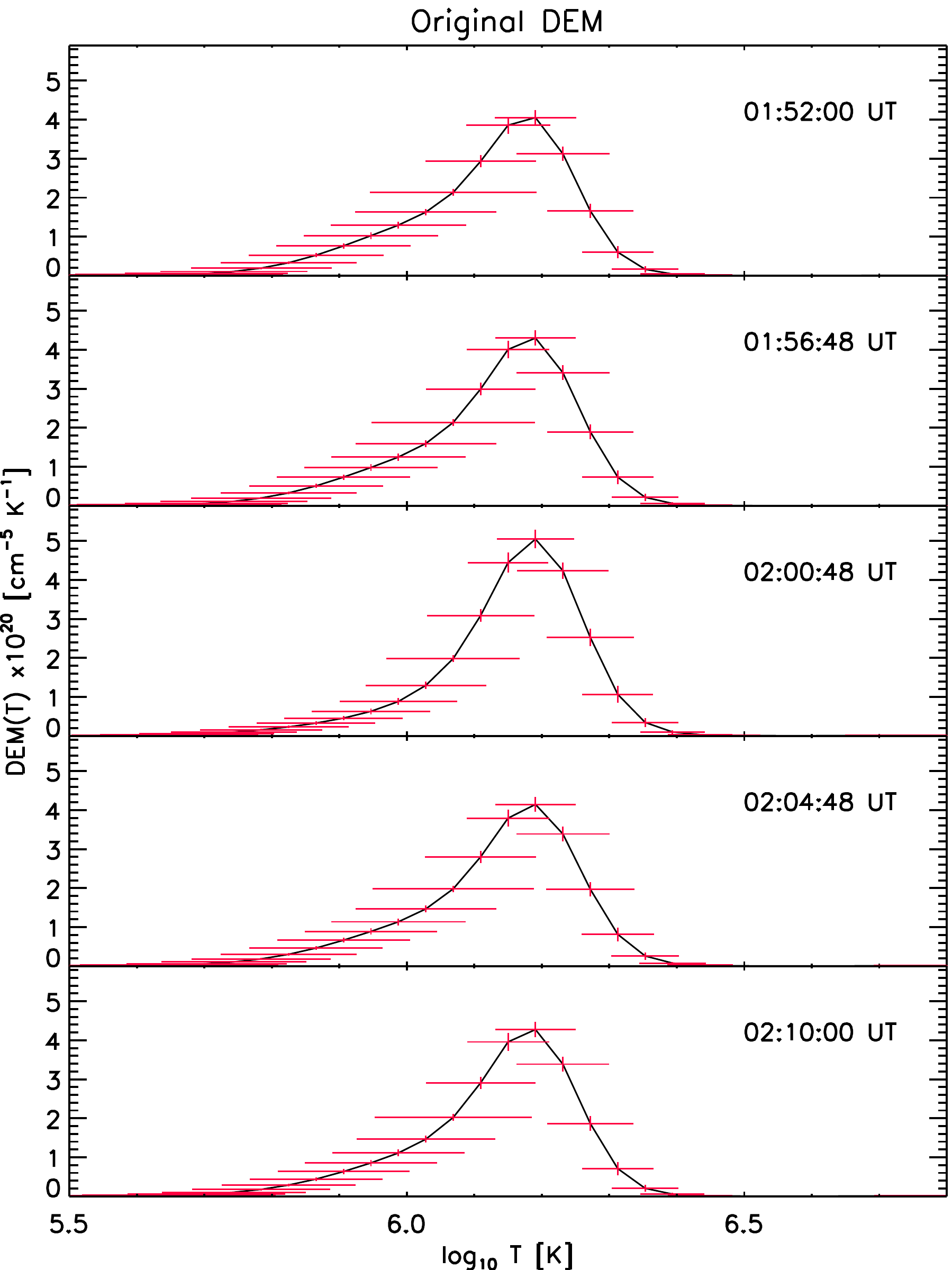}
\hspace{0cm}
\includegraphics[width=0.45\textwidth]{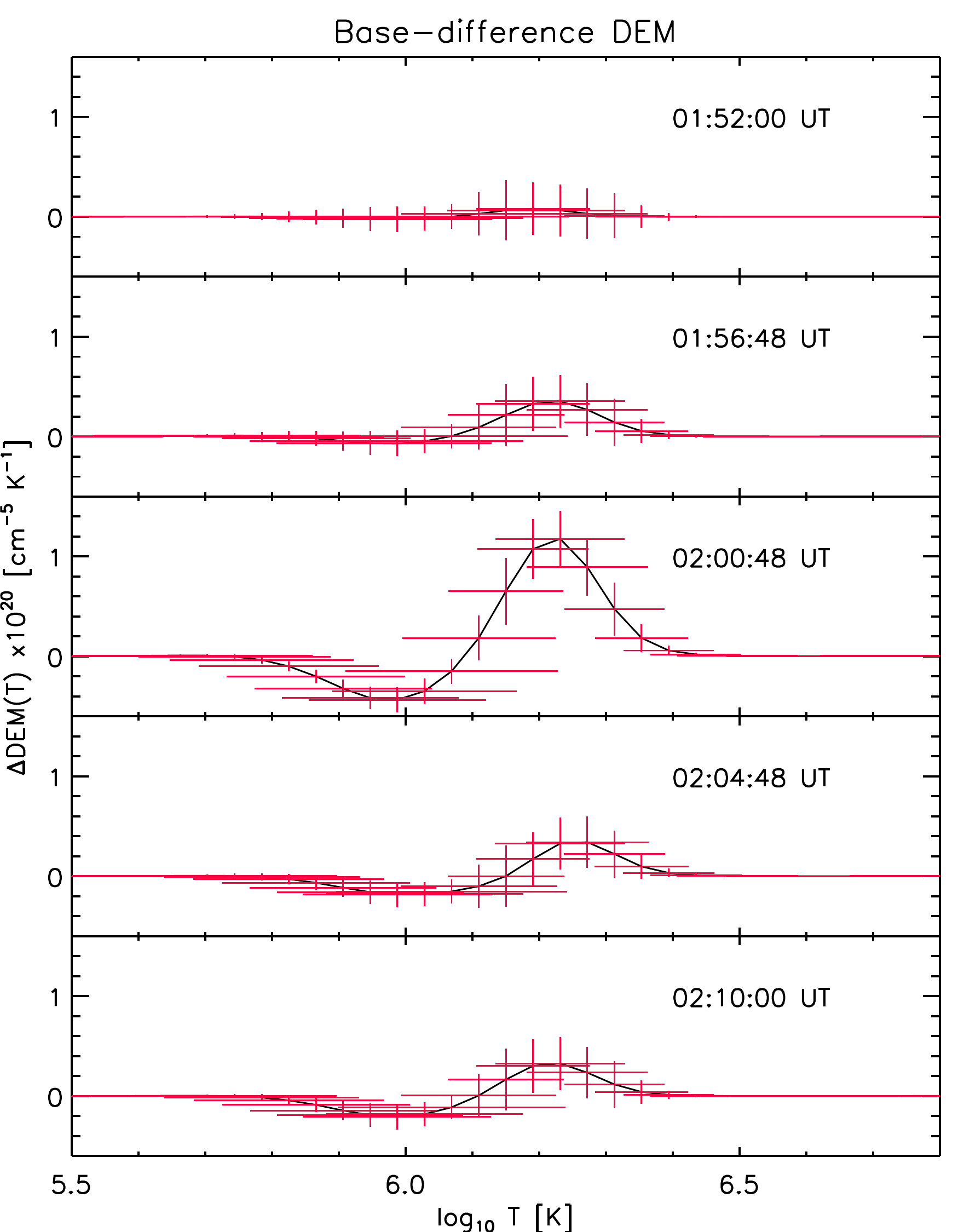}
\caption{DEM curves from the original images \textbf{(left)} and $\Delta$DEM obtained by subtracting the base DEM (at 01:45\,UT) from the original data DEMs \textbf{(right)} for ROI55 at the time steps indicated in each of the panels. Same as the time steps given in Figure~\ref{profiles}.}
\label{dem}
\end{figure*}
In order to study the response of the corona to the passage of the EUV wave along our chosen directions of study, we selected regions of interest (ROIs), hereafter referred to as ROI55 and ROI210 corresponding to 55$\degree$ and 210$\degree$ from the solar west, respectively, (marked by green asterisks in Figure~\ref{sector}) and studied the plasma properties of these regions as a function of time. The size of ROI55 was chosen such that at a given time step (01:58:48\,UT for this direction) the whole wave pulse occupies the ROI. This size was found to be $\approx$168\,Mm. The same could not be done for ROI210 due to the presence of a stationary brightening which is discussed later in Section~\ref{210direction}.

Figure~\ref{kinematics} shows the EUV wave kinematics in the AIA~211\,\AA\ channel along the two sectors covering the ROIs. The position of the EUV wave (indicated by the blue arrows in Figures~\ref{profiles}~and~\ref{profile210}) was determined from the perturbation profiles of the 211\,\AA\ channel, as the wave is clearly distinguishable here, and the same position was used in the remaining channels.
 We identified the leading front by fitting a Gaussian function to the wave pulse. The MPFIT routine was used to calculate the Gaussian parameters along with its errors. The leading front, $x_{\text{lead}}$, was defined as
\begin{equation}
x_{\text{lead}}={p_{1}}+2{p_{2}},
\end{equation}
where $p_{1}$ is the centroid of the Gaussian and $p_{2}$ is the width.
Accordingly, the 1$\sigma$ errors in $x_{\text{lead}}$ is given by
\begin{equation}
\Delta x_{\text{lead}}=\pm \left(\Delta{p_{1}}+2\Delta{p_{2}}\right).
\end{equation}
The speed of the propagating wave front thus estimated, in the 211\,\AA\ channel, along the 55$\degree$ direction was found to be 437$\pm$ 19\,\kms\ and along the 210$\degree$ direction was 366$\pm$ 44\,\kms.

\subsection{Study along 55$\degree$ direction}
\label{55direction}

In Figure~\ref{profiles} we show the perturbation profiles from AIA 211\,\AA\ channel along 55$\degree$ direction at five different time steps corresponding to before (panels 1, 2), during (panel 3) and after (panels 4, 5) the passage of the EUV wave from ROI55, marked in red. The average flux from this region taken from all the six AIA coronal channels were used as inputs for the DEM analysis.

The curves resulting from our DEM analysis are shown in Figure~\ref{dem}. The left panels correspond to DEM curves derived from the perturbation profiles constructed from the original data. The right panels are $\Delta$DEM curves derived by subtracting the DEM of the base image (taken at 01:45\,UT) from the DEMs shown in the left panels. Here, the negative values imply a reduction in emission with respect to the pre-event corona.  These curves are plotted for the same time steps as the perturbation profiles shown in Figure~\ref{profiles}. The DEMs in the left panels were used for further calculations while the $\Delta$DEMs in the right panels illustrate the temperatures at which there is a reduction/increase in emission compared to the pre-event corona. These changes are not very obvious from the original data DEMs (left panels).

In Figure~\ref{dem}, the top two rows show the DEM and $\Delta$DEM before the EUV wave arrived at ROI55. The right panels show very small change in DEM values and is comparable to the background emission. The middle row corresponds to the DEM and $\Delta$DEM when the EUV wave pulse was at ROI55 while the bottom two rows of this figure shows the DEM and $\Delta$DEM when most/all of the wave pulse has moved out of ROI55.

\begin{figure}[ht]
\hspace{-0.5cm}
\includegraphics[width=8.7cm]{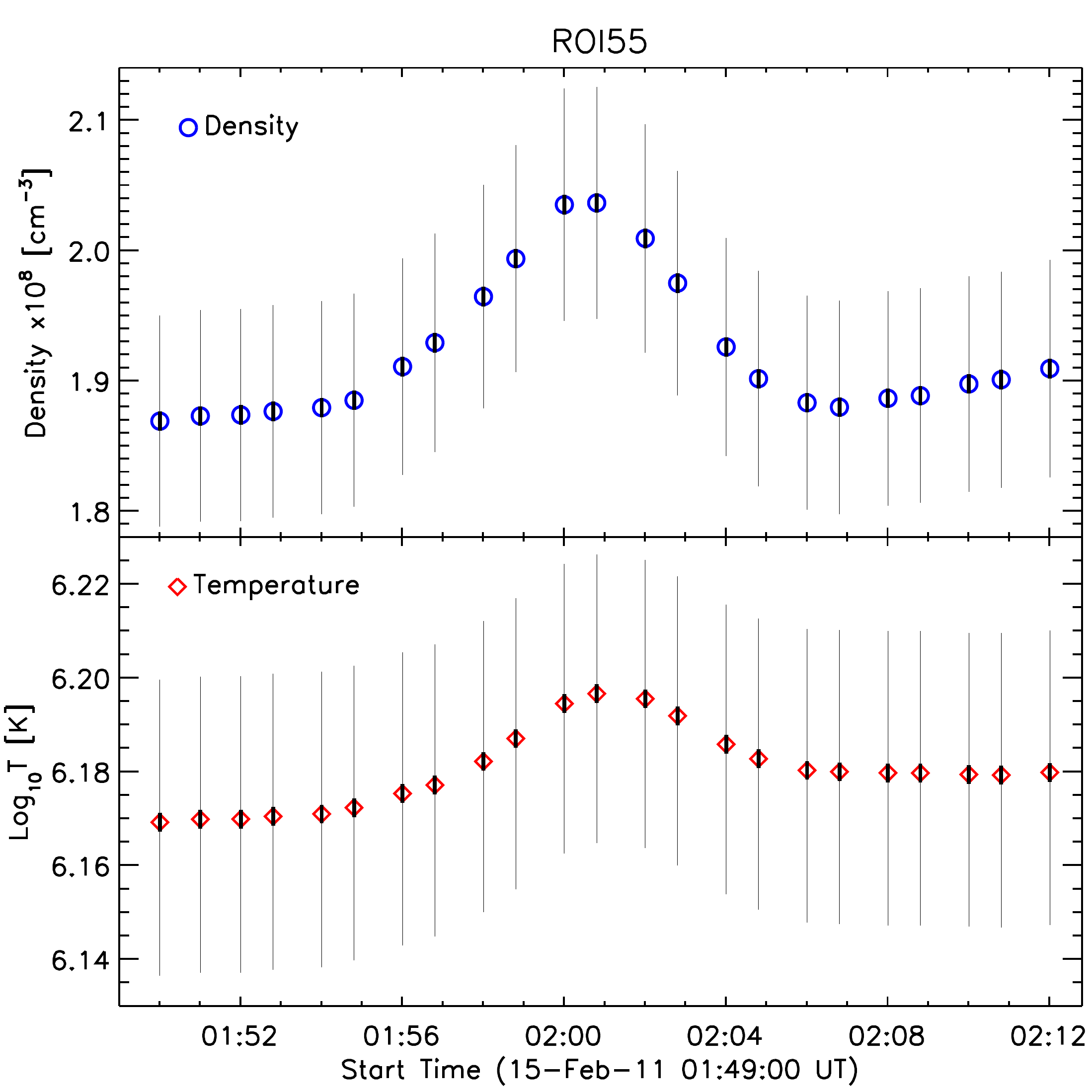}
\caption{Average density \textbf{(top)} and temperature \textbf{(bottom)} derived from the DEM evolution at ROI55 as a function of time. Thin black lines represent the absolute errors while the thick shorter lines show the relative error.}
\label{density}
\end{figure}

Using equations \ref{temp_eq} and \ref{dem_eq} we derived the averaged temperatures and densities, respectively, from the DEMs obtained with the original data (left panels of Figure~\ref{dem}). Figure~\ref{density} shows the time evolution of the derived quantities along with its errors. The temperatures and densities obtained in the time range of 01:50\,UT and 01:53\,UT are approximately constant as the EUV wave had not yet reached ROI55. We took the average of these quantities as the background temperature and density. From the values plotted in Figure~\ref{density} we measured an 8.7\% increase in density compared to the background (from 1.87$\times10^{8}$\,cm$^{-3}$ to 2.03$\times10^{8}$\,cm$^{-3}$) as the EUV wave passed through ROI55. A similar calculation for temperature showed a 6.3\% increase from 1.48\,MK to 1.57\,MK.

After the EUV wave left ROI55 we saw that at first the density falls back to the background levels (at 02:06:48\,UT in Figure~\ref{density}) and then it increases slightly until 02:12:00\,UT. However, during this time the temperature is not seen to reach the background levels and it remains at an elevated level.
\begin{figure}[h]
\includegraphics[width=8.5cm]{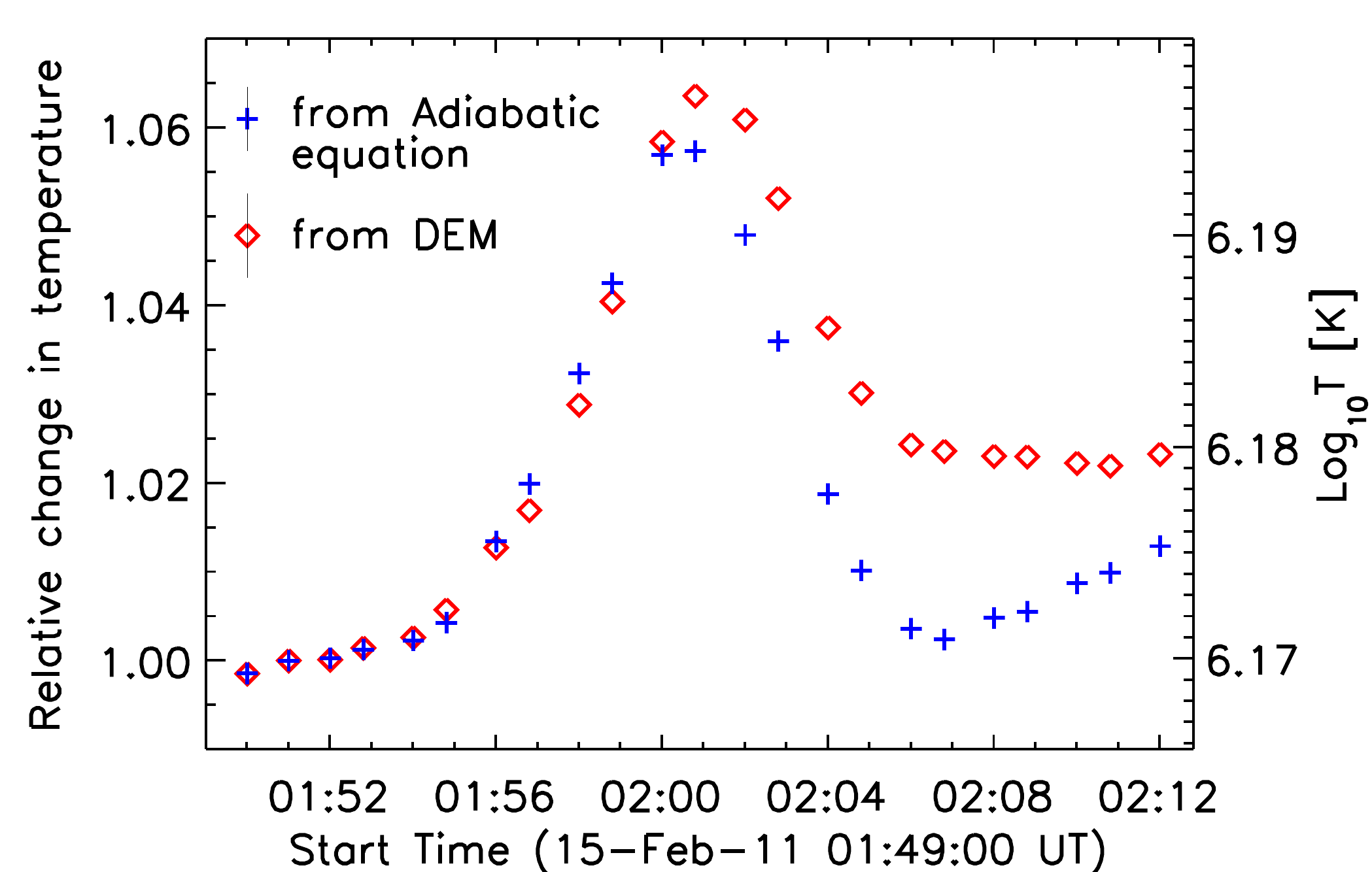}
\caption{The expected change in temperature for an adiabatic process are plotted as blue plus symbols and the temperature changes obtained from the DEM diagnostics are plotted as red diamonds. The absolute temperature scale is plotted on the right axis.  For ease of viewing the errors are shown in the legend rather than on the data points.}
\label{adiabatic}
\end{figure}

We calculated the absolute errors in density to be of the order of 0.16$\times10^8$\,cm$^{-3}$ and those in log$_{10}$T to be of the order of 0.06 dex.  We note here that the main contribution to the error in density is from the error in the value of $h$ which is $\approx$10\%. In this paper we discuss about the relative changes and not the absolute values. Hence, we want to determine errors which are more relevant to the current study rather than using the absolute errors. We have shown that the temperature of the event studied here varies between log$_{10}$T=6.17 and 6.19. This falls within the range where the AIA channels have very good temperature sensitivity. So it is more appropriate to discuss errors for changes within this specific temperature range rather than for the whole. Synthetic Gaussian DEM curves were folded through the AIA instrument response and the result was used as input to the HK code to obtain DEM reconstructions. The mean temperature and total emission measure calculated using these curves were used to derive the errors. A detailed description of the method is given in the appendix. The errors obtained, using this method, in density is of the order of 0.014$\times$10$^8$\,cm$^{-3}$ and in temperature is 0.004\,dex. Figure~\ref{density} shows for comparison both, the absolute errors and errors derived using equations~\ref{EMerr} and \ref{Terr}. In the subsequent images we show only the newly derived errors that relate to the relative changes in density and temperature.
 
To verify the theory of adiabatic compression of the plasma at the wave front we used the adiabatic relation between temperature and density
\vspace{-0.2cm}
\begin{equation}
\frac{T}{T_{0}}=\left(\frac{\rho}{\rho_{0}}\right)^{\gamma-1}
\label{adiabatic_eq}
\end{equation}
 (assuming the adiabatic index $\gamma=5/3$ for fully ionized plasma). We calculated the change in temperature $(T/T_0$) from our DEM estimated values and compared them with the expected change in temperature during an adiabatic process as derived from the density changes $(\rho/\rho_0$) using equation~\ref{adiabatic_eq} (see Figure~\ref{adiabatic}). We found that during the heating phase the observed and expected values are in good agreement but the same is not true for the cooling phase. We also note that the secondary slow increase in density from 02:06:48\,UT to 02:12:00\,UT is not resulting in an adiabatic increase of temperature.

\begin{figure}[ht]
 \vspace{-0.2cm}
\includegraphics[width=8.5cm]{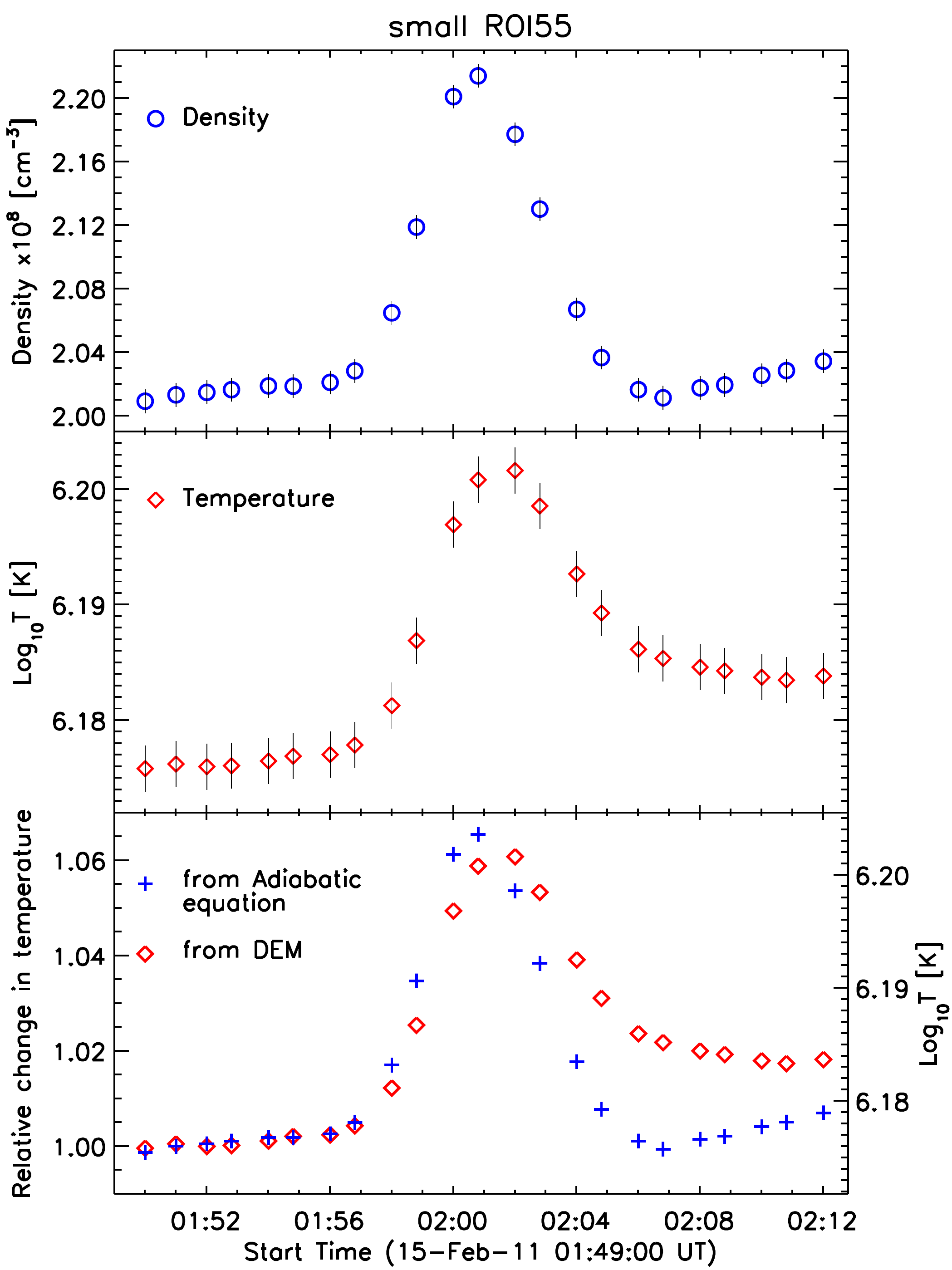}
\caption{Sample plots for test done with finer spatial binning at ROI55. Average density \textbf{(top)} and temperature \textbf{(middle)} derived from the DEM evolution as a function of time. \textbf{Bottom:} The expected change in temperature for an adiabatic process are plotted as blue plus symbols and the temperature changes obtained from the DEM diagnostics are plotted as red diamonds. The absolute temperature scale is plotted on the right axis. For ease of viewing the errors are shown in the legend rather than on the data points.}
\label{smallroi}
\end{figure}

 We did the same analysis as described above by decreasing the size of ROI55 to 96\,Mm. In this case of finer spatial binning we see that the general evolution is the same but the derived changes are larger: a 9.9\% increase in density and a 6.1\% temperature increase with respect to the background values (see Figure~\ref{smallroi}). Again, the heating phase is consistent with adiabatic compression at the wave front while the cooling phase is not.
 
\subsection{Study along 210$\degree$ direction}
\label{210direction}
\begin{figure}[ht!]
\includegraphics[width=8.5cm]{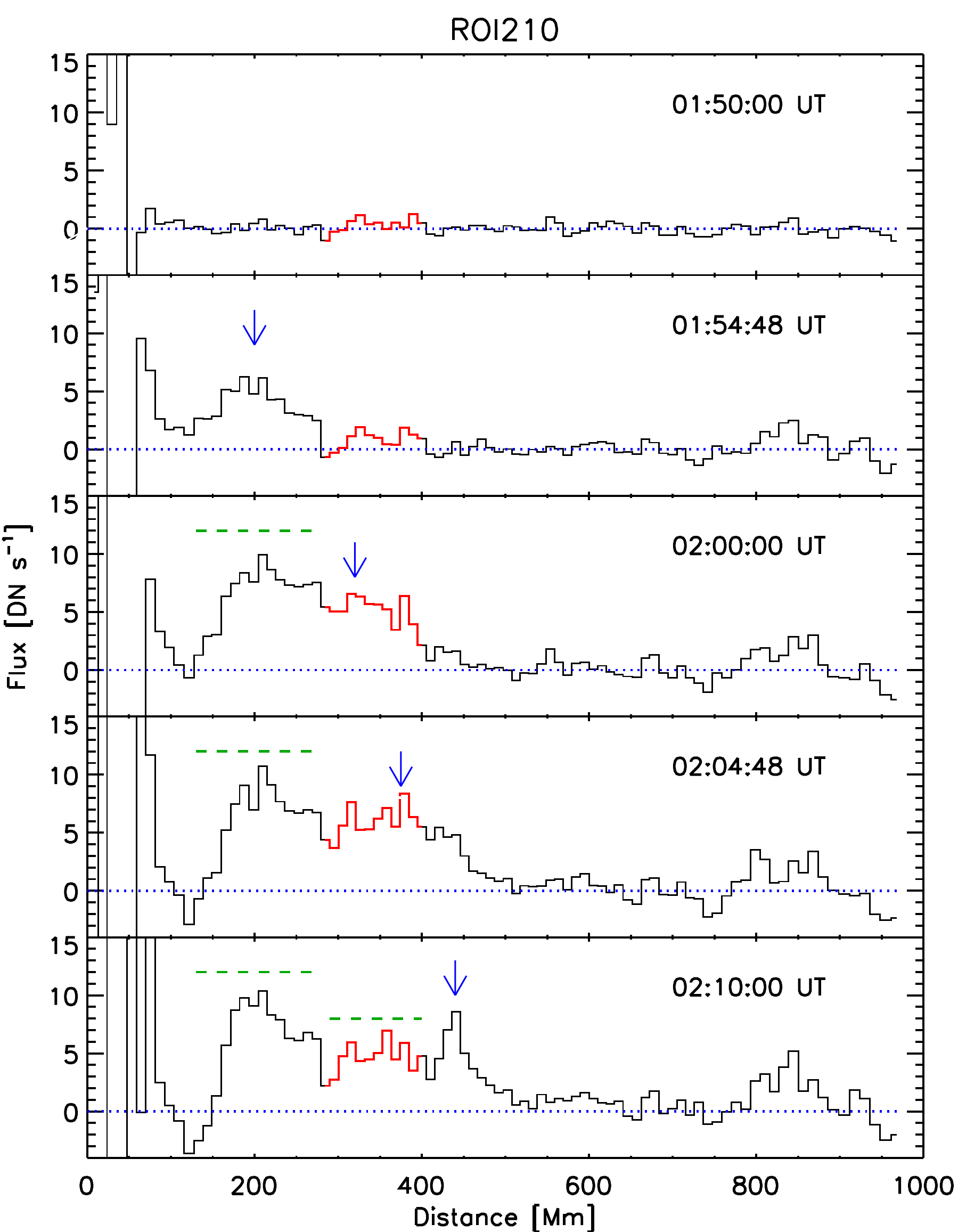}
\caption{Perturbation profiles along the 210$\degree$ direction derived from AIA 211\,\AA\ base-difference images at five different time steps as indicated in the panels. The blue arrow points to the propagating EUV wave. The section of the plots indicated in red corresponds to ROI210 marked by green asterisks along the $210\degree$ direction in Figure~\ref{sector}. The dashed green horizontal line marks the position of the stationary brightening.}
\label{profile210}
\end{figure}

For comparison we performed a similar study along the 210$\degree$ direction. We show example plots of perturbation profiles in Figure~\ref{profile210} with the position of the EUV wave indicated by the blue arrows. In this direction the clear identification of the EUV wave was complicated by the presence of a stationary brightening close to the location of the flare. The position of the brightening is indicated by a horizontal green dashed line in the panels of Figure~\ref{profile210} where it appears. Such brightenings have been previously reported for different events \citep{Delannee1999, Attrill2007, Cohen2009, Muhr2011}.

 From  Figure~\ref{profile210} we see that the brightening does not exist prior to the passage of the EUV wave (top panel). By comparing panel 2 with panels 3, 4 and 5 we can say that the brightening occupies the same position as the first appearance of the EUV wave pulse and continues to remain there at least for the next 15\,min. In panel 5 of this figure we see that a secondary stationary brightening is formed adjacent to the first one after the EUV wave pulse has moved forward. The clear distinction of the wave pulse from the stationary brightening is difficult. We selected ROI210 such that it does not include the primary stationary brightening.

\begin{figure}[ht]
\includegraphics[width=8.7cm]{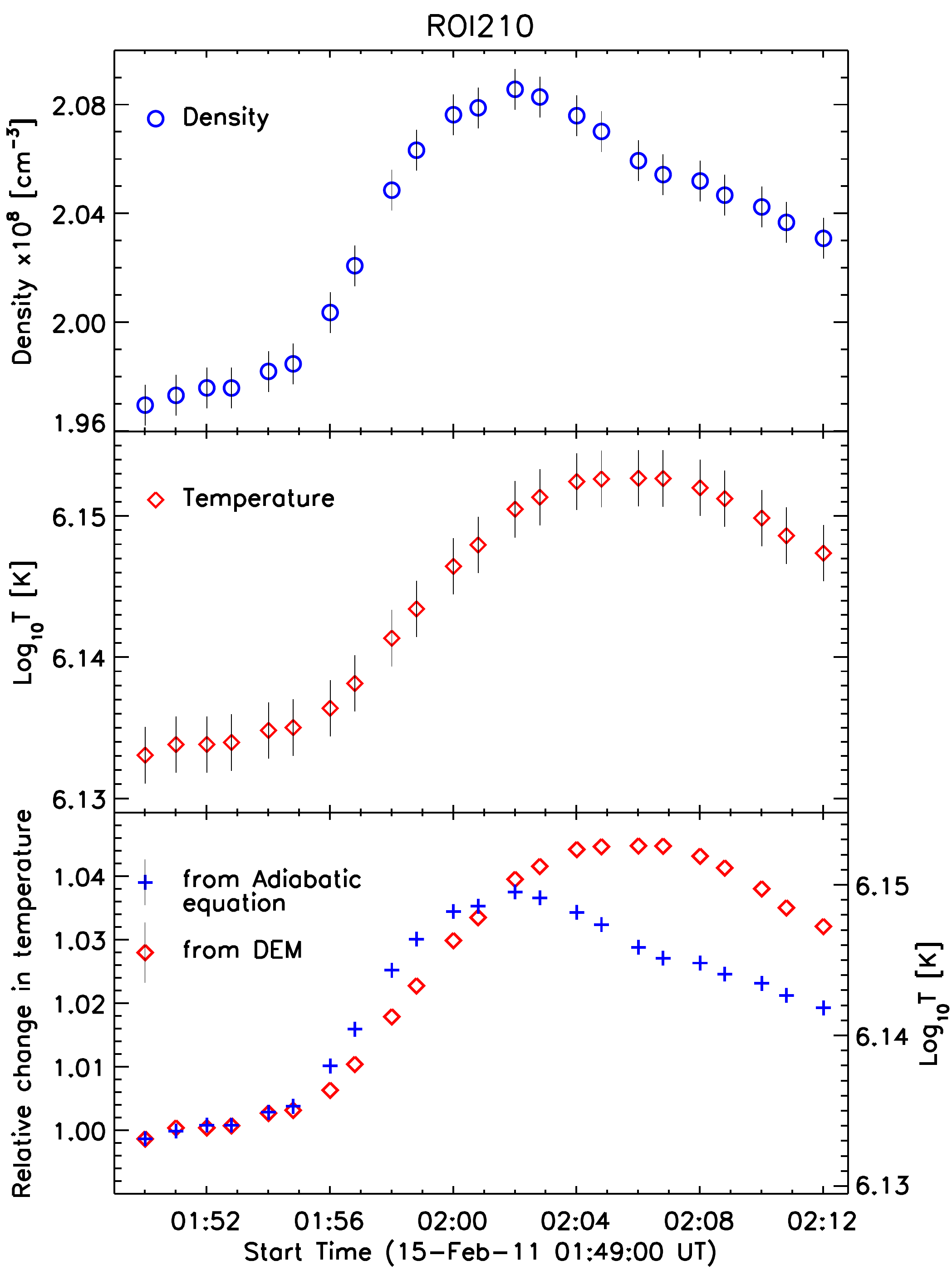}
\caption{Same as Figure~\ref{smallroi} but for ROI210.}
\label{density210}
\end{figure}

  We followed the same procedure as described in Section~\ref{55direction} and constructed DEM curves from the original data for ROI210. We used the DEMs to determine densities and temperatures and studied the time evolution of these quantities (see Figure~\ref{density210}).

From Figure~\ref{density210} we found that the relation between density and temperature is different from what we observe in ROI55. The density in ROI210 peaks at 02:02\,UT, the temperature continues to rise for 4\,mins after this time until 02:06\,UT before it begins to fall. From the values plotted in Figure~\ref{density210} we calculated a 5.7\% increase in density and a 4.5\% increase in temperature. The density increased from 1.97$\times10^{8}$\,cm$^{-3}$ to 2.08$\times10^{8}$\,cm$^{-3}$ and the temperature increased from 1.36\,MK to 1.42\,MK.

We also calculated the adiabatic changes for ROI210 and found it to be similar to ROI55 with one exemption. From Figure~\ref{density210} (bottom) we see that the temperature rise phase until 02:02\,UT nicely fit with the adiabatic relationship, but the observed increase thereafter, 02:02\,UT to 02:06\,UT, is not consistent with heating due to adiabatic compression and requires other additional agents.


\section{Discussions}
\label{discussions}
The appearance of the EUV wave as a dark front in the 171\AA\ channel has often been interpreted as emission decrease in this channel due to heating. 
In our study with the $\Delta$DEM curves we have shown that, when the EUV wave passes over the ROI there is an increase in the DEM values in the region bounded by 6.1~$\leq$~log$_{10}$\,T~$\leq$~6.4 (which coincides with the peak of the temperature response curves for the AIA 193\,\AA\ and 211\,\AA\ channels) while there is a decrease in the DEM in the region of 5.8~$\leq$~log$_{10}$\,T~$\leq$~6.1 (corresponding to the peak temperature of the AIA 171\,\AA\ filter). 
From our observations the average pre-event temperature T$_{0}$~=~1.48\,MK (see Figure~\ref{density}) and lies to the right of the response peak of the 171\,\AA\ channel (along the negative gradient of the response curve) and to the left of the response peaks of the 193\,\AA\ and 211\,\AA\ channels (along the positive gradient of the response curve). This implies that the mean DEM-weighted temperature is shifting locally along the response curve, $R(T)$, and the change in emission is proportional to the derivative of the response curve $\mathrm{d}R(T)/\mathrm{d}T$. A temperature increase from T$_{0}$ means it is moving down along the 171\,\AA\ response curve and up along the 193\,\AA\ and 211\,\AA\ response curves, implying a decrease in flux in the 171\,\AA\ channel (darkening) and an increase in flux in the 193\,\AA\ and 211\,\AA\ channels (brightening). This confirms that the dark front observed in 171\,\AA\ channel is a result of plasma being heated to higher temperatures.
 After the EUV wave has passed the ROI, we notice that the peaks of the $\Delta$DEM values are diminished and have shifted slightly towards lower temperatures indicating that there is cooling of the plasma behind the wave.

In the relaxation phase at ROI55, we have pointed out that the density falls back to background levels. In case of a coronal dimming behind the wave we would expect the density to drop below the background values. This observation suggests that the EUV wave is not followed by a dimming region or in other words, this is a wave phenomenon which only displaces the plasma in its current position and does not carry it forward.

\begin{figure}[ht]
\includegraphics[width=8.5cm]{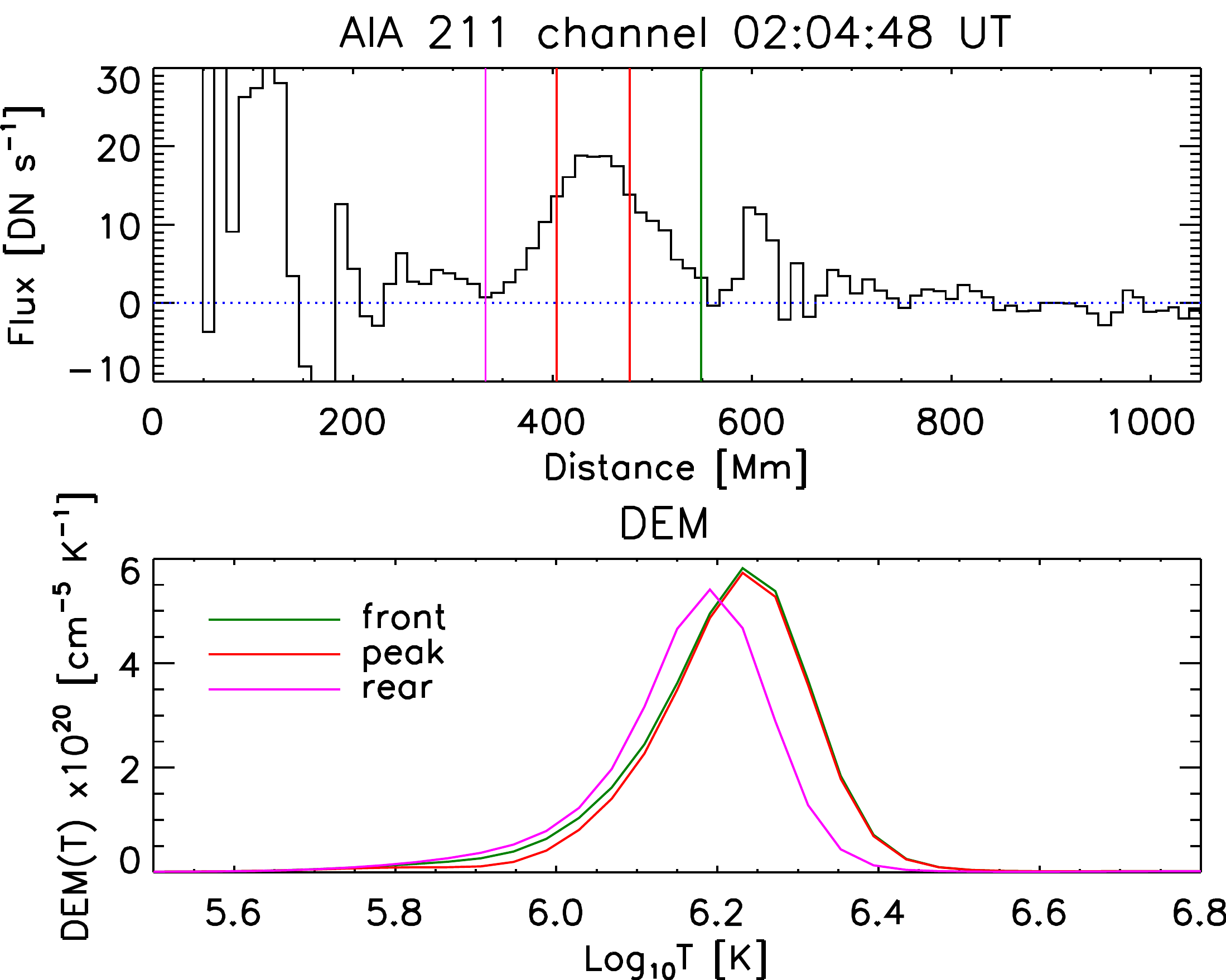}
\caption{{\bf Top:} Perturbation profile from the 211\,\AA\ channel taken at 02:04:48\,UT. The vertical lines divide the wave pulse into three segments: front, peak and rear (from right to left). {\bf Bottom:} The DEMs for the three segments shown above and as indicated by the legend.}
\label{pulse}
\end{figure}

 We have also studied in detail the plasma characteristics across the wave pulse by dividing it into three segments corresponding to front, peak and rear as shown in Figure~\ref{pulse}. We found that for a particular time step the segments corresponding to the peak and front of the pulse were hotter and denser than the rear. This sequence was true for all the time steps where a clear identification of the wave pulse was possible. Average temperatures for the front, peak and rear segments of the wave pulse were 1.75$\pm$0.008\,MK, 1.77$\pm$0.008\,MK and 1.60$\pm$0.007\,MK, respectively. Such an observation indicates that heating due to compression is taking place at the wave front which is consistent with the behavior of a compressive wave.

The relaxation phase in ROI55 shows that the temperature remains at an elevated level while the density returns to the background levels. This deviation could be due to non-linear effects at play for a such a global wave pulse, which are not accounted for in equation~\ref{adiabatic_eq}. The peak density amplitude derived is 6\%, which is actually quite big considering the assumption of small perturbations (linear theory) from which the standard magneto-acoustic waves are derived. Alternatively, it could be a result of local energy release triggered by the passing wave front, e.g. reconnection with favorably oriented field lines \citep{Attrill2007}, which provides heating in addition to the adiabatic process. 
There have been several examples (as mentioned in Section~\ref{intro}) of EUV waves disturbing structures that are present along its path. When the EUV wave passes over the quiet-Sun, it disturbs the local magnetic field lines which can result in magnetic reconnection behind the wave front. Although reconnection is not the primary mechanism producing the wave front, it can be a side effect of the passing wave. Such occurrences will contribute in increasing the local temperature while it need not necessarily change the local density.

Based on the relative emission changes at the wave front in the AIA 171\,\AA, 193\,\AA\ and 211\,\AA\ filters, \cite{Schrijver2011} estimated that the observations are consistent with adiabatic compression for plasma temperatures between 1.2 and 1.8\,MK. Assuming adiabatic compression as the only source of heating, they have shown that the observed intensities in the three filters are consistent with adiabatic compression at the wave front for T=1.3\,MK. Assuming isothermality, they derived a maximum density increase of 10\% associated with a temperature increase of 7\% for the east-west propagation of the EUV wave over the quiet-Sun. They interpreted these enhancements as a result of compression of the expanding CME against the stalks of the overlying helmet-streamer.

We have taken a different approach and have performed a DEM analysis to study the plasma properties at the wave front. Our study is a detailed quantitative analysis of the compression and relaxation phase, at and behind the wave front, respectively. The increase in density and temperature of  9\% and 6\%, respectively, in the 55\degree direction and 6\% and 5\%, respectively in the 210\degree direction that we have calculated in the current study are in basic agreement with the results of Schrijver et al. (2011). However, we interpret this phenomenon as a compressive fast-mode wave  for the following reasons: (1) we have found that the evolution of temperature and density are consistent with adiabatic compression, as is expected in the case of a compressive wave; (2) we are able to detect a wave pulse quite close to the active region while in the streamer scenario the increase in intensity should only be visible when the CME approaches the streamer footpoints; (3) we have analyzed different segments across the wave pulse and found that heating and compression is taking place close to the front of the wave which is the likely scenario for a compressive wave whereas in the current shell model \citetext{as suggested in \citealt{Schrijver2011}}, the highest temperature is expected to be close to the rear part of the pulse which is adjacent to the current layer.

The stationary brightening that we detect in the 210$\degree$ direction has been previously interpreted as compression of plasma near the foot points of opening field lines due to the expansion of CME flanks \citep{Delannee1999, Muhr2011} or as ongoing reconnection between expanding flux ropes and surrounding favorably oriented magnetic field \citep{Attrill2007, Cohen2009}. In either case this could result in energy release and associated temperature increase. We see the effect of this when we compare the densities and temperatures at ROI210. Firstly, the density in ROI210 does not return to the background level as it did in ROI55. Secondly, there is an increase in temperature even after the density has reached its peak. These observations can be attributed to the presence of the secondary stationary brightening which occurs at ROI210 (see panel 5 of Figure~\ref{profile210}). We also confirm this from the comparison of actual temperature change with the expected adiabatic change. The temperature increase from 02:02\,UT until 02:06\,UT is not due to adiabatic compression. 

In our study we show that the theory of adiabatic compression at the wave front is consistent for both directions that we have studied, with the exclusion of the effect of the stationary brightening in the 210$\degree$ direction. However, the relaxation phase is more difficult to account for. Since the EUV wave alters structures on its path, it cannot be expected that a particular region remains the same before and after the passage of the wave. Deviation from the adiabatic relationship in the relaxation phase could be attributed to local changes triggered by the passage of the EUV wave that provide additional release of energy through a process which is not adiabatic.

\section{Summary and Conclusion}
\label{conclusion}
We present here a study on the EUV wave associated with the large X-class flare of 2011 February 15. We studied the response of the quiet corona to the passage of the EUV wave using DEM analysis. We selected two different directions to study the wave. One was ``undisturbed'' by external factors (at 55$\degree$) while the second was ``disturbed'' by the presence of a stationary brightening (at 210$\degree$).

In the 55$\degree$ direction, we have shown that the wave front approaching ROI55 compresses the plasma underneath resulting in heating that is purely adiabatic in nature. There is a 6\% increase in temperature corresponding to a 9\% increase in density.  When the EUV wave departs from ROI55 we see that the density falls back to values close to its initial state suggesting that there is no dimming region behind the EUV wave. This observation supports that the phenomenon is a wave only displacing plasma in its current position. On the other hand, the temperature drops slowly and does not return to its background values. It remains at an elevated level after some time.

In the 210$\degree$ direction, the approaching EUV wave front at first compresses the plasma adiabatically. 5\% increase in temperature is associated with 6\% increase in density. Later a stationary brightening is formed at the position of ROI210 which affects the density and temperature of the local plasma. We notice that when the EUV wave has departed ROI210 the temperature continues to rise briefly despite the density decrease indicating that the wave passage has induced some local energy release processes. Unlike ROI55, the density in this region does not return to its initial state.

In this study, it has been quantitatively shown that the process of adiabatic compression is heating the coronal plasma at the wave front, consistent with a MHD fast mode wave. The behavior of the EUV wave during the heating phase, as we have shown here, provides support to the wave nature of  the phenomenon under study. The temperature structure across a wave pulse shows heating is taking place at the front of the wave and this provides additional evidence of the event being a fast magnetosonic wave.



\acknowledgments

KV, AMV, and KD acknowledge the Austrian Science Fund (FWF): P24092-N16. MSM is funded by the Leverhulme trust. IGH acknowledges support from a Royal Society University Research Fellowship. EPK recognises financial support from the STFC Consolidated Grant. KV and AMV thank ISSI for the support of the team ``The Nature of Coronal Bright Fronts''. KV is grateful to Jaroslav Dud\'{\i}k for his help. The AIA data are courtesy of SDO (NASA).





\appendix
\section{}
\begin{figure}[ht]
\includegraphics[width=8cm]{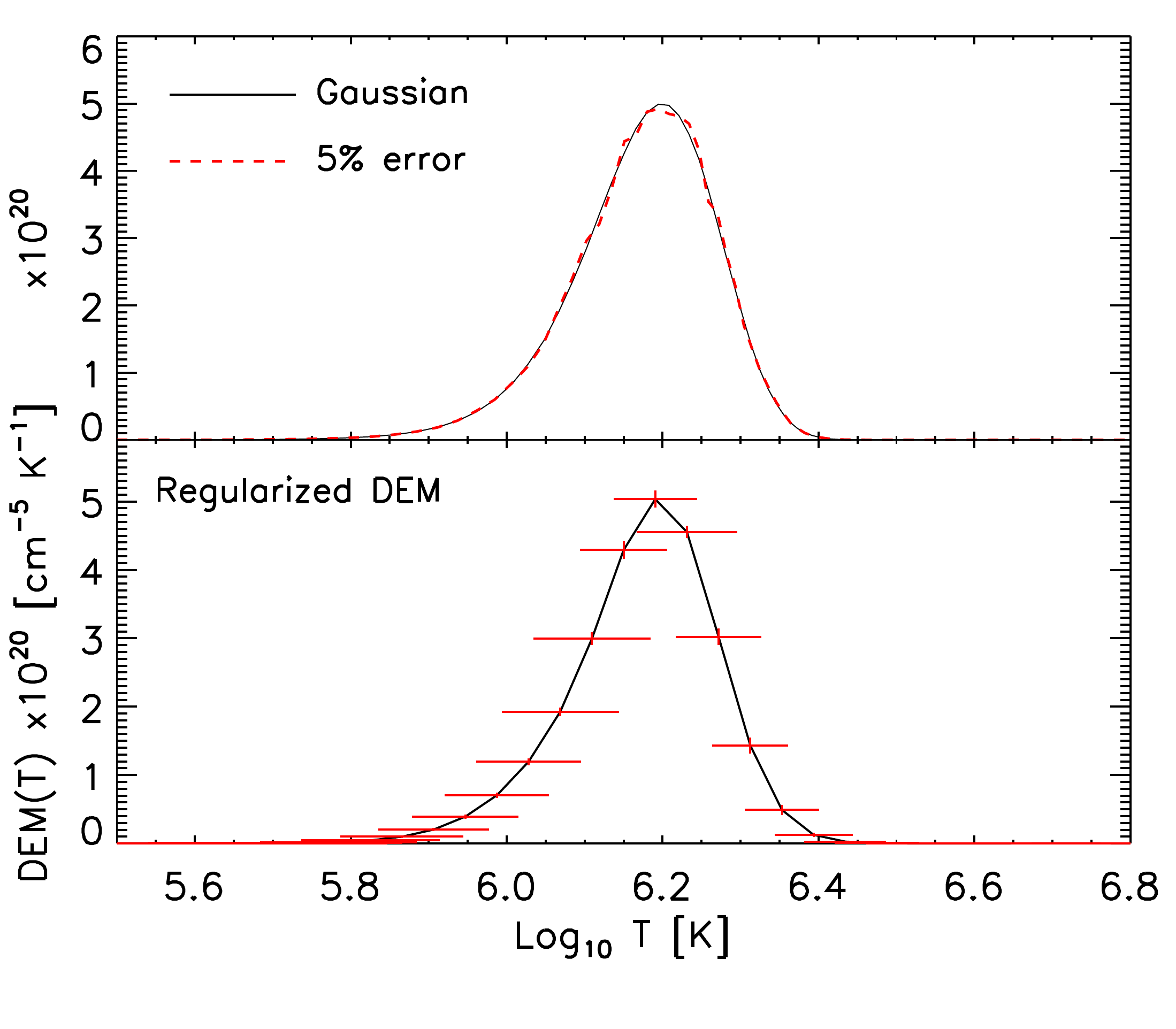}
\vspace{-0.5cm}
\caption{An example of an input Gaussian curve along with 5\% error {\bf(top)} and its corresponding DEM curve {\bf(bottom)}. }
\label{gauss_dem}
\end{figure}

\begin{figure}[ht]
\includegraphics[width=8cm]{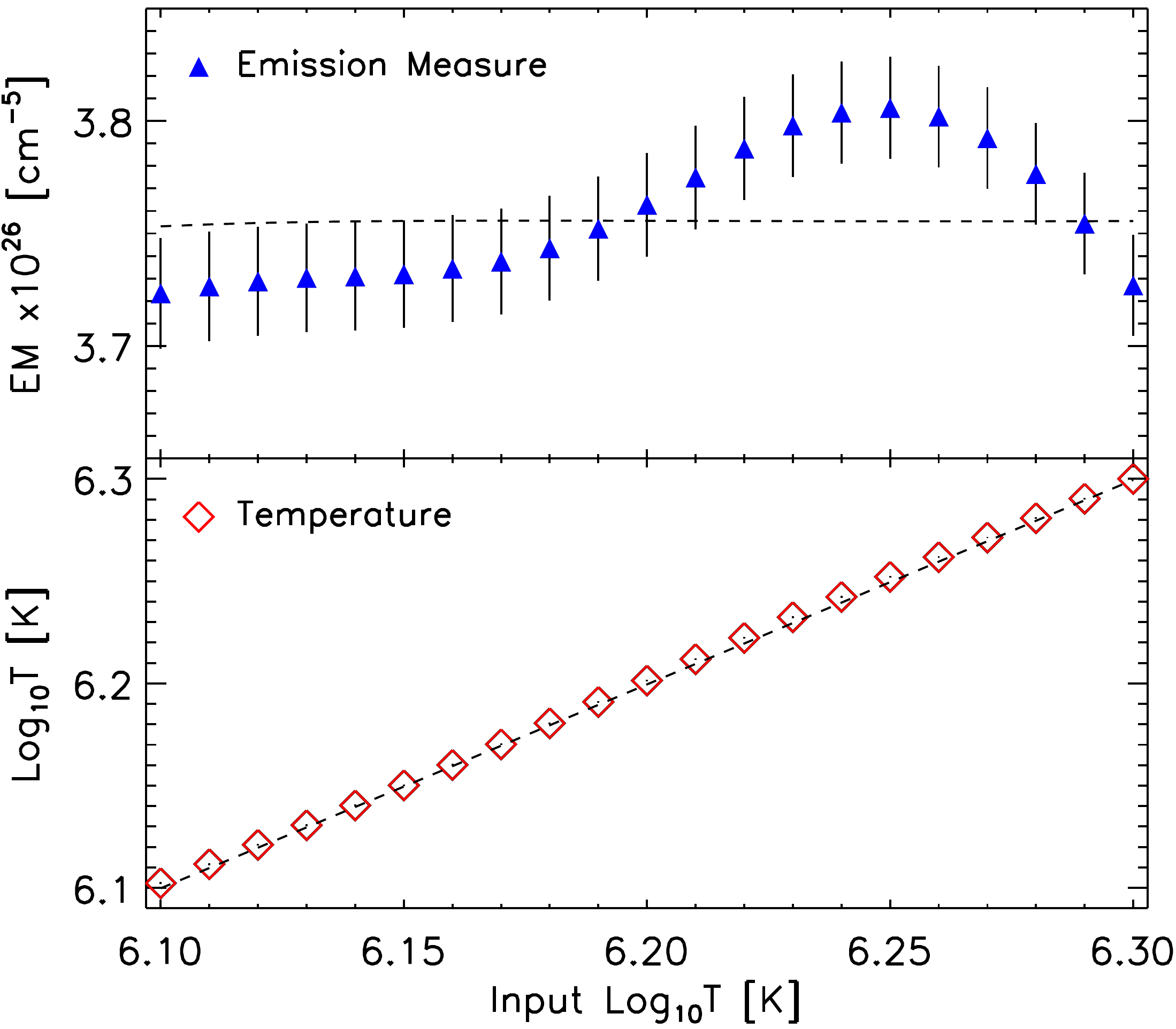}
\caption{The emission measure {\bf(top)} and temperature {\bf(bottom)} as calculated from the HK code. The dashed lines in each of the plots depicts the values calculated from the input Gaussian.}
\label{em_temp}
\end{figure}

\vspace{-0.5cm}
We have used the following method to test the sensitivity of the HK code that is used in this study and to derive errors within the temperature range under consideration here. A typical DEM curve is most similar to a Gaussian curve. So we have used a Gaussian, drawn as a function of temperature, as a proxy for a DEM. The amplitude and width of the Gaussian is taken to be close to the observed DEMs. Since the temperature of the event studied in this paper varies from log$_{10}$T=6.17 to 6.19 for the 55\degree direction and log$_{10}$T=6.13 to 6.15 in the 210\degree direction, we varied the position of the center of the Gaussian peak from log$_{10}$T=6.10 to 6.30 in intervals of 0.01. The temperature sensitivity of all the AIA EUV channels are the best within the temperature range of the event studied in this paper. So we use this method to derive errors that is most relevant to this study, where we are mostly interested in the {\it changes} in plasma temperature and density (rather than the accuracy of the absolute values). Using the relation
\begin{equation}
I_{filter}=\Sigma R_{filter}(T)\,f(T)\,\Delta T,
\end{equation}
where $R_{filter}(T)$ corresponds to the temperature response function of an EUV filter in AIA and $f(T)$ is the input Gaussian representing a DEM, we were able to obtain the simulated counts for the six EUV channels. These counts were further used as inputs for the HK code to obtain the corresponding DEMs. A comparison between the input Gaussian and the DEM from HK code is shown in Figure~\ref{gauss_dem}.
From these DEM curves, the weighted average temperature and emission measure ($EM(T)$) were calculated using equations\,\ref{temp_eq} and \ref{em_eq}, respectively and plotted as a function of the Gaussian peak temperature:
\begin{equation}
\label{em_eq}
EM(T)=\int \phi(T)\,\mathrm{d}T.
\end{equation}
This exercise was repeated by introducing 3 different levels of random errors (3\%, 5\% and 10\%) to the input Gaussian curves that represent DEMs. The results from introducing 5\% error is shown in Figure~\ref{gauss_dem}. For each level of error applied to the input Gaussian, 100 realizations of randomness were performed. From the 100 realizations, the mean and standard deviation of the emission measure and temperature were calculated for each error level. These plots for the 5\% random error are shown in Figure~\ref{em_temp}. 
We notice that the HK code is able to reproduce the Gaussian input temperatures quite accurately with a standard deviation of 0.0004 dex for 5\% error while the standard deviation for emission measure is 2.33$\times10^{24}$\,cm$^{-3}$ in the considered range of Gaussian peak temperature (log$_{10}$T=6.10 to 6.30). In this study we are interested in the changes in the plasma parameters so we want to derive errors which are more relevant than just the absolute errors. We use the following relations to define the typical errors in temperature and emission measure within the temperature range under study here:
\begin{equation}
\label{EMerr}
\Delta EM~=~\overline{\lvert EM_{DEM}-EM_{input}\rvert},
\end{equation}

\begin{equation}
\label{Terr}
\Delta log\,T~=~\overline{\lvert log\,T_{DEM}-log\,T_{input}\rvert},
\end{equation}
where $T_{DEM}$, $EM_{DEM}$ correspond to the values plotted in Figure~\ref{em_temp} and $T_{input}$, $EM_{input}$ correspond to the dashed lines shown in the same figure. Using this method we obtain errors in temperature as 0.004 dex and in emission measure as 5.2$\times10^{24}$ which corresponds to an error of  0.014$\times10^8$\,cm$^{-3}$ in density. These were used instead of the absolute errors since they are more specific to the temperature range associated with the event we are studying. Comparison between the two errors is shown in Figure~\ref{density}.




\bibliography{Vanninathan.bbl}

\clearpage



\clearpage

\end{document}